\documentclass[prl,aps,amssymb,twocolumn,epsfig]{revtex4}
\usepackage{epsfig}
\usepackage{amsmath}
\usepackage{amssymb}
\usepackage{bm}
\usepackage{dcolumn}% Align table columns on decimal point

\newcommand{\beq}{\begin{equation}}
\newcommand{\eeq}{\end{equation}}
\newcommand{\bea}{\begin{eqnarray}}
\newcommand{\eea}{\end{eqnarray}}

\begin{document}

\title{$S_3$ Quantum Hall Wavefunctions}

\author{Steven H. Simon${}^{1}$, Edward H. Rezayi${}^{2}$, and Nicolas Regnault${}^3$ \\ } \address{${}^1$Rudolf Peierls Centre for Theoretical Physics, 1 Keble Road, Oxford, OX1 3NP, UK \\${}^2$Department of Physics, California State University, Los Angeles, California 90032 \\ ${}^{3}$Laboratoire Pierre Aigrain,
ENS, CNRS, 24 rue Lhomond, 75005 Paris, France}
\begin{abstract}
We construct a family of quantum Hall Hamiltonians whose ground states, at least for small system sizes, give correlators of the $S_3$ conformal field theories.   The ground states are considered as trial wavefunctions for quantum Hall effect  of bosons at filling fraction $\nu=3/4$ interacting either via delta function interaction or delta function plus dipole interaction.  While the $S_3$ theories can be either unitary or nonunitary, we find high overlaps with exact diagonalizations only for the nonunitary case, suggesting that these wavefunctions may correspond to critical points, possibly analogous to the previously studied Gaffnian wavefunction.  These wavefunctions give an explicit example which cannot be fully characterized by their thin-torus limit or by their pattern of zeros.
\end{abstract}
\date{November 9, 2009}
\pacs{
71.10.Pm %      Fermions in reduced dimensions (anyons, composite fermions, Luttinger liquid, etc.) (for anyon mechanism in superconductors, see 74.20.Mn)
%73.43.-f        % Quantum Hall effects
73.43.Cd        % Theory and modeling
73.21.Ac        % Multilayers
}
\maketitle

One of the major breakthroughs in the theory of quantum Hall effect was the realization of the close correspondence between quantum Hall wavefunctions and conformal field theories (CFTs).  This correspondence suggested the possibility that quasiparticle excitations of certain quantum Hall states might have nontrivial (``nonabelian") braiding statistics\cite{MooreRead} --- a property that, if true, could be useful for error resistant quantum information processing\cite{NayakRMP}.   While this CFT correspondence has been extremely powerful, only a very few nontrivial CFTs have successfully been used to generate reasonable trial wavefunctions\cite{MooreRead,ReadRezayi,Nonspinpolarized}  (we define ``success" in a moment).   In fact, among spin-polarized single-component wavefunctions  (which we will focus on throughout this paper\cite{Nonspinpolarized}), it appears that the only successes of this approach have been the Read-Rezayi\cite{ReadRezayi} series including the Moore-Read\cite{MooreRead} state and the Laughlin state.  While many other wavefunctions have been proposed\cite{Gaffnian,Haffnian,Estienne,Bernevig,BernevigSimon,Fuchs}, serious problems plague these attempts:  (1) CFT approaches that do not produce an explicit wavefunction\cite{Estienne,Fuchs} are difficult to study.  (2) Of the new explicit wavefunctions that have been proposed, many do not correspond to rational unitary CFTs\cite{Bernevig,BernevigSimon,Gaffnian,Haffnian}, and there is increasingly strong evidence\cite{Read} that only rational unitary CFTs can describe a gapped phase of matter.   (Although other CFTs may describe interesting critical points between gapped phases, and may therefore be worth studying nonetheless).   (3) With the exception of the above-mentioned successes, the non-unitary Gaffnian\cite{Gaffnian}, and the nonrational Haffnian\cite{Haffnian}, no one has found an explicit Hamiltonian whose ground state is one of these proposed CFTs.  (4) even should one propose a Hamiltonian that produces a valid unitary CFT wavefunction as its ground state, there is still a serious issue of whether this phase of matter can be realized in any reasonable experiment.   We define ``success" of a wavefunction by these four criteria.  In passing, we note two other partially successful  wavefunction constructions in Refs.~\onlinecite{Bonderson} and Ref.~\onlinecite{Hansson} which both fail condition (3).

In the current paper we propose a family of wavefunctions based on the so-called $S_3$ CFTs\cite{S3} which describe bosons at $\nu=3/4$ (or fermions at $\nu=3/7$).  While far from declaring these wavefunctions to be ``successful" on the scale of the Read-Rezayi series\cite{ReadRezayi}, our results are nonetheless favorable with respect to the above listed criteria.  In particular, we develop a family of Hamiltonians that, at least for small systems, generates a family of $S_3$ CFT wavefunctions which includes both unitary and nonunitary cases.  We find that these trial wavefunctions can have very high overlap with wavefunctions of exact diagonalizations of potentially realistic Hamiltonians corresponding to  rotating Bose gases\cite{Cooper}.   Interestingly, we find that the high overlaps coincide with nonunitary $S_3$ wavefunctions.  This behavior, reminiscent of the previously studied Gaffnian wavefunction\cite{Gaffnian}, suggests criticality in both cases.   We believe these results may shed some important light on the general applicability of nonunitary CFTs to quantum Hall physics in general.  Further, our results suggest that wavefunctions based on the higher generation parafermion CFTs\cite{HigherGen,BernevigSimon,Estienne}  may generally be of experimental and theoretical interest.

The $S_3$ wavefunctions are within the larger class of generalized parafermion (or $W$-algebra) wavefunctions\cite{BernevigSimon,Estienne,HigherGen} that generalize the parafermionic Read-Rezayi states\cite{ReadRezayi}.  It is useful to start by reviewing some of the properties of the $S_3$ CFTs\cite{S3} before describing the wavefunctions that can be built from them.

The family of $S_3$ CFTs have two simple currents $\psi$ and $\psi^\dagger$ with conformal dimension $h=4/3$  (these are analogs of $\psi_1$ and $\psi_2$ of the $\mathbb{Z}_3$ parafermion theory\cite{S3,HigherGen}).  These fields satisfy the operator product expansions (OPEs)
\begin{eqnarray}
 \psi(z_1) \psi(z_2) &=& \lambda \, (z_{1}-z_{2})^{-4/3} \,\,  \psi^\dagger(z_2)+ \ldots  \\
     \psi^\dagger(z_1) \psi^\dagger(z_2) &=& \lambda \, (z_{1}- z_{2})^{-4/3} \,\, \psi(z_2)+  \ldots  \\
     \psi(z_1) \psi^\dagger(z_2) &=& (z_{1}-z_{2})^{-8/3} I(z_2)  +  \ldots \end{eqnarray}
where $I$ is the identity field,  $c$ is the central charge, and $\ldots$ represents terms less singular by integer powers of $(z_1 - z_2)$.   The constant $\lambda$ is related to the central charge by\cite{S3}
$
    \lambda^2 = 4(8-c)/(9c)
$.  Within this family of CFTs there exists a series of rational minimal models, which we denote $S_3(p,p')$  having corresponding central charge
$ c = 2 \left(1 - \frac{3 (p - p')^2}{4 p p'} \right)$ for $p$ and $p'$ positive integers. Note that there is a rational minimal CFT from this family arbitrarily close to any $c \leq 2$, although the only unitary members of this set occur for the discrete series $p'=p+4 \geq 7$.  See Table 1 for several examples of such minimal models.

We will focus on the case of quantum Hall effect of bosons, although the fermionic wavefunctions can also be considered quite analogously (as we will see below, the bosonic case seems to be of potential experimental interest).  Following the general approach for constructing quantum Hall wavefunctions from Refs \onlinecite{MooreRead,ReadRezayi} we write a wavefunction for bosons at filling fraction $\nu=3/4$ as
\begin{equation}
\label{eq:wavefunction}
\Psi = \langle \psi(z_1) \ldots \psi(z_N) \rangle \prod_{i<j} (z_i - z_j)^{4/3}
\end{equation}
where the number of particles $N$ is taken to be divisible by 3.
The full wavefunction also includes a measure $\prod_{i=1}^N \mu_i$ which we do not write explicitly.  For a planar geometry the measure is
$
%\label{eq:plane}
\mu_i = \exp(-|z_i|^2/4 \ell_0^2)
$
with $\ell_0$ the magnetic length, whereas for a spherical geometry\cite{ReadRezayi}
$\mu_i = (1 + |z_i|^2/4 R^2)^{-(1 + N_{\phi}/2)}
$
where $R$ is the radius of the sphere, and $N_{\phi}$ is the total number of flux quanta through the sphere.   Note that on the sphere, this wavefunction occurs for flux related to the number of particles by $N_{\phi} = \frac{4}{3} N - 4$  (i.e., the shift is 4).

From the OPEs it is easy to see that $\Psi$ does not vanish when three particles come to the same position, but vanishes as four powers as the fourth particle arrives at this position (This generalizes the $\mathbb{Z}_3$ Read-Rezayi state which vanishes as two powers as the fourth particle arrives).   As the four particles come together, the wavefunction must vanish proportional to some fourth degree translationally invariant symmetric polynomial.  As pointed out in Ref.~\onlinecite{Pseudo}, there is a two dimensional space of such polynomials.
Let us parameterize this with orthonormal vectors
\begin{eqnarray}
    |\varphi_\theta^0\rangle &=& \cos \theta |\varphi_1\rangle + \sin \theta |\varphi_2\rangle \\
        |\varphi_\theta^\perp\rangle &=& -\sin \theta |\varphi_1\rangle + \cos \theta |\varphi_2\rangle \\
\end{eqnarray}
where $
\varphi_1 = c_1\mbox{$\sum_{1 \leq i < j \leq 4}$} \, (z_i - z_j)^4  $
and $\varphi_2$ is chosen orthogonal to $\varphi_1$ with respect to the measure $\mu_i$.   On the plane
$ \varphi_2 = c_2 (z_1 + z_2 + z_3 - 3 z_4) \times \mbox{cyclic}
$ where $c_1$ and $c_2$ are normalizations, such that  $\langle \varphi_i | \varphi_j \rangle = \delta_{ij}$ for  $i=1,2$ so that
  $  \langle \varphi_\theta^k | \varphi_\theta^m \rangle = \delta_{km}
$ where  $k,m = 0, \perp$.

We now define a Hamiltonian as outlined in Ref.~\onlinecite{Pseudo}, $H=\tilde H + H_\theta$ where $\tilde H$ gives positive energy to any four particles having relative angular momentum less than four and
\begin{equation}
 H_\theta= \sum_{i ,j , k,l}  |\varphi^0_\theta(z_i,z_j,z_k,z_l)\rangle \,\langle \varphi^0_\theta(z_i,z_j,z_k,z_l)|
\end{equation}
forces any four particles to approach each other proportional to $\varphi^\perp_\theta$ (or as a higher degree polynomial) or else they will pay an energy penalty.  Thus, this Hamiltonian is projecting the four particle behavior to be $\varphi^\perp_\theta$.   Using the approach of Ref.~\onlinecite{Pseudo} this Hamiltonian may be written either in terms of four-particle pseudopotentials, or in terms of derivatives of a four-particle delta-function.   While we have no proof that our Hamiltonian will result in a unique ground state generally, we find numerically that for up to 15 bosons on a sphere, the ground state (for $N_\phi=\frac{4}{3} N - 4$ with $N$ divisible by 3) is indeed unique, and therefore must correspond to the CFT generated wavefunction Eq.~\ref{eq:wavefunction} for the appropriate central charge corresponding to the chosen $\theta$.  (Furthermore it is found that the ground state wavefunctions generally satisfy the product rule discovered in Ref.~\onlinecite{Product}). To identify the central charge associated with a given $H_\theta$  consider the limits
\begin{eqnarray}
\label{eq:G}
    G &=& \lim_{z \rightarrow 0}  \left(z^{-4} \left[\lim_{z_4 \rightarrow z_3=z} \left[\lim_{z_2 \rightarrow z_1=0} \Psi \right] \right]\right) \\
    \label{eq:F}
     F&=& \lim_{z_4 \rightarrow z \rightarrow 0} \left(z^{-4} \left[\lim_{z_3  \rightarrow 0} \left[\lim_{z_2 \rightarrow z_1=0} \Psi \right] \right] \right)
\end{eqnarray}
From the OPEs it is easy to show that
\begin{equation}
\label{eq:GF}
  G/F = \lambda^2  = 4 (8-c)/(9c)
\end{equation}
Thus, taking the analogous limits for the four particle wavefunction $\varphi_\theta^\perp$ (which gives the limiting four particle behavior of the ground state) one can easily determine $c$ for any given $\theta$.  We are thus able to numerically generate the $S_3$ quantum Hall wavefunctions (Eq. \ref{eq:wavefunction}) corresponding to any central charge.
For certain values of the central charge, the generated wavefunction results in wavefunctions previously proposed (See Table 1).   Note that generating the ground state wavefunction does not address a host of questions, such as whether the same Hamiltonian will produce a unique ground state for $N > 15$ (although this seems likely), whether this Hamiltonian is gapped, what the excitations look like, or what the spectrum is when additional flux quanta are added\cite{endnotespectrum}.

\begin{table}
\begin{tabular}{|c c|c|c|}
\hline
\multicolumn{2}{|c|}{Unitary Theories}  & $c$ & $\theta/\pi$   \\
\hline $S_3(3,7)$ & tricritical potts & 6/7 &   0.288    \\
     $S_3(4,8)$ & $\mathbb{Z}_6$ parafermion &   5/4 &  0.378 \\
     $S_3(5,9)$  &   &   22/15 & 0.420   \\
     $S_3(6,10)$  & $[\mathbb{Z}_3 \, \mbox{parafermion}]^2$  &   8/5 &  0.443  \\
     & \vdots & \vdots &  \vdots     \\
      \multicolumn{2}{|c|}{accumulation point} &  2 & 0.5     \\
      \hline
\multicolumn{2}{|c|}{NonUnitary Rational Examples}  & $c$ & $\theta/\pi$
 \\ \hline
$S_3(2,7)$  & & -19/28 & -0.0113   \\
       $S_3(3,10)$  & & -9/20 &  0.0176    \\
       $S_3(1,7)$  &  Jack  &  -40/7 &  -0.188   \\
    $S_3(1,3)$   &  &   0   &  0.0913         \\
       \hline
  \multicolumn{2}{|c|}{Other Examples}  & $c$ & $\theta/\pi$
 \\ \hline
 &  & 8  &  -0.307                \\
 &  & -4 &   -0.166 \\
 &  &  -1   &  -0.0443   \\
 &   &  -0.584$\ldots$ & 0     \\  \hline
\end{tabular}
\caption{Examples of $S_3$ wavefunctions, $c$ is the central charge and $\theta$ is the tuning parameter of the Hamiltonian.  Values of $\theta/\pi$ are approximate (except 0 and 0.5) and are calculated for the $N=12$ system size.  The $\mathbb{Z}_6$ case condenses the $\psi_2$ field as discussed originally in Ref. \onlinecite{ReadRezayi} and developed further in Ref.~\onlinecite{Pattern}. The Jack wavefunction is the $(k,r)=(3,4)$ case from Ref. \onlinecite{Bernevig}.  The $c=8$ case is the $\nu=3/4$ wavefunction from Ref.~\onlinecite{Gaffnian}, and the $c=-4$ is from Ref.~\onlinecite{Regnaultwf}.  The accumulation point with $c=2$ is the symmetrized product of three $\nu=1/4$ Laughlin wavefunctions.}
\end{table}

The generated wavefunctions, since they do not vanish when 3 particles come to the same point but vanish as four powers when the fourth arrives, are examples of cluster state wavefunctions as discussed for example in Refs.~\cite{Bernevig,SimonProjection,Pattern,Torus}.  However, we emphasize again that this statement does not by itself fully determine the wavefunction --- one needs to specify $\varphi_\theta^\perp$, the precise manner in which the wavefunction vanishes.  This means that neither the thin torus limit\cite{Torus} (or root state\cite{Bernevig} of the full clustered polynomial fractional quantum Hall wavefunction\cite{SimonProjection,Bernevig}), nor the pattern of zeros\cite{Pattern} uniquely specify the wavefunction.  Thus, our current example starkly points out the weaknesses of several proposed schemes for characterizing quantum Hall wavefunctions in general.

We now turn to the question of whether these wavefunctions could have applicability to experimentally realistic situations.  We choose to look at quantum Hall effect of bosons, which may be relevant to rotating cold Bose gases\cite{Cooper}.   Assuming delta function interactions, Ref.~\onlinecite{Chang} showed compellingly that both filling fractions $\nu=1/2$ (where the ground state is exactly the Laughlin wavefunction) and $\nu=2/3$ are well described by composite fermion physics (meaning both the ground state and excited states are well described).  However, the next member of the composite fermion series, $\nu=3/4$, while still a gapped state\cite{endnote2}, fits much less well to the composite fermion description\cite{Chang}. This was one of the reasons we began to seek an alternate wavefunction for this state.  Note that the composite fermion wavefunction and our $S_3$ wavefunctions compete with each other directly being that on the sphere they both occur for $N_\phi = \frac{4}{3} N - 4$.    Within this work we also consider altering the delta function interaction by adding an additional dipolar interaction as well\cite{DipoleBosons,Cooper} which also could be experimentally relevant.  We quantify the amount of dipole interaction we have added by specifying the so-called Haldane pseudopotential coefficient ratio $V_2/V_0$.

\begin{figure}
\begin{center}
\includegraphics[width=0.8\columnwidth]{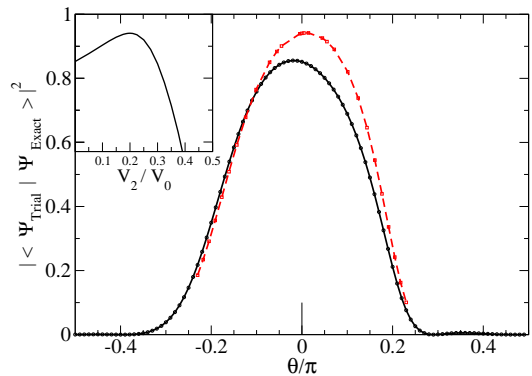}
\end{center}
\caption{Squared overlap of $N=12$ bosons at $\nu=3/4$ exact wavefunction with the $S_3$ trial wavefunction as  a function of the tuning parameter $\theta$.  Solid curve shows the overlap of the trial wavefunctions with the exact ground state of bosons interacting via repulsive delta function interaction.   Dashed curve is the overlap with the exact ground state of bosons interacting via repulsive delta function plus dipole interaction where the amount of dipole interaction is set such that $V_2/V_0 = 0.2$.  Inset:  Overlap at $\theta=0$ as a function of amount of dipole interaction $V_2/V_0$.  The vertical axis of the inset is aligned with the vertical axis of the main plot.}
%\label{fig:hund}
\end{figure}

In Fig. 1, we show the overlap of our $S_3$ trial wavefunctions with the exact ground state for $N=12$ bosons at $N_\phi = \frac{4}{3} N - 4 = 12$, as a function of the tuning parameter $\theta$ in the Hamiltonian.  The solid line is the overlap with the exact ground state of bosons interacting via delta function interactions.  The $L=0$ Hilbert space of this system has 127 dimensions, so the high overlaps of over 85\%  should be considered to be significant.  When we include dipolar interaction terms as well, the overlap increases further, as shown in the inset, reaching over 94\% near $V_2/V_0 = 0.2$.  This increase of overlap with added dipole interaction is reminiscent of the behavior of bosons at $\nu=1$ where\cite{Regnault} for similar sized systems, the overlap of the Moore-Read wavefunction with the exact ground state at $V_2/V_0=0$ is about 88\%, increases to about $95\%$ with increasing $V_2/V_0$ and then collapses above $V_2/V_0 \approx 0.2$.  In comparison, the overlap here of the composite fermion wavefunction with the exact ground state\cite{Chang} is only about 74\% at $V_2/V_0=0$. then decreases monotonically with increasing $V_2/V_0$.   The dashed line in the main plot shows the overlap of our trial wavefunctions with the exact ground state of bosons interacting via delta function plus dipole interaction with $V_2/V_0=0.2$.

The maximum overlaps in these diagonalizations occur near $\theta=0$ which corresponds to a wavefunction which vanishes proportional to $\varphi^{\perp}_0 = \varphi_2$ when four particles come together.   For $N=12$ as shown in the figure, $\theta=0$ corresponds to central charge $c\approx-.584$.  (In the thermodynamic limit
$\theta=0$
corresponds to $c=-8/11$).       Numerically, the maximum overlap with the ground state of delta function interaction (solid line) occurs at $c\approx-0.739$ whereas for the case of delta function plus dipole with $V_2/V_0 = 0.2$ (dashed plot) the maximum occurs for $c\approx-0.503$.     At any rate, these negative values of the central charge indicate that such a CFT must necessarily be nonunitary.    The entire unitary series occurs for $.25 < \theta/\pi \leq 0.5$ and the overlaps with the exact ground states are low.  In searching for a CFT that corresponds to our data, it is worth noting that we can always find a rational (but not typically unitary) CFT arbitrarily close to any desired central charge with $c < 2$ by choosing appropriate $p$ and $p'$ in $S_3(p,p')$.  However, if we insist that $p$ and $p'$ are both ``reasonably small" (say 10 or less), then the only theories that gives $-1 < c < 0$ are $S_3(2,7)$ and $S_3(3,10)$ (See Table 1).

Our results are reminiscent of the physics of the Gaffnian\cite{Gaffnian}($\nu=2/3$ for bosons) in many ways.  In both cases we find very high overlaps with exact diagonalization, despite being nonunitary theories.  In both cases, there is a competing composite fermion wavefunction at the same flux which also has high overlap with the exact ground state.  Our $S_3$ wavefunction vanishes roughly proportional to $\varphi_2 = [(z_1 + z_2 + z_3 - 3 z_4) \times \mbox{cyclic}]$ when four particles come together, whereas the Gaffnian vanishes analogously as $[(z_1 + z_2 - 2 z_3) \times \mbox{cyclic}]$ when three particles come together.   This analogy suggests that the entire composite fermion series will have similar behavior and that the $k^{th}$ member of the composite fermion series ($\nu=k/(k+1)$ for bosons, or $k/(2k+1)$ for fermions) will compete similarly with a generalized parafermion wavefunction\cite{BernevigSimon,Estienne,HigherGen} with $\mathbb{Z}_k$ symmetry.

Our current understanding of the Gaffnian suggests that\cite{Gaffnian}, as a nonunitary CFT, it is actually a critical point between two phases --- one of which is the composite fermion phase, and the other (less well understood) phase perhaps being some sort of strong pairing phase.  One can explicitly tune through this critical point by varying the two particle $V_0$ interaction: for positive $V_0$ the wavefunction has increasingly high overlap with the composite fermion wavefunctions, whereas for negative values of $V_0$ the wavefunction rapidly obtains very low overlaps.  A calculation of wavefunction fidelity (to be published) suggests the critical point is at $V_0=0$.  Further, one may conjecture that the high overlaps between the Gaffnian and the composite fermion wavefunction at the same filling fraction is a sign that the typical composite fermion wavefunctions are somehow ``close" to this critical point.  We conjecture that there may be similar physics occurring for this case of $\nu=3/4$  bosons (or $\nu=3/7$ fermions) where the critical theory here is one of the $S_3$ wavefunctions. Indeed, the behavior with an added $V_0$ interaction appears to mimic that of the Gaffnian quite closely: positive $V_0$ leaves the wavefunction relatively stable with high composite fermion overlap, whereas negative $V_0$ pushes the wavefunction to a different phase which has very low overlap with composite fermions.  We conjecture that the entire composite fermion series follows this pattern.

As a summary, let us now revisit our above criteria for a successful wavefunction.  (1) While we have not written an analytic wavefunction, we have nonetheless numerically generated the wavefunction corresponding to the $S_3$ CFTs at least for small systems.  (2) At least some members of this family are unitary CFTs (3) We have written explicit Hamiltonians for which these wavefunctions are the unique ground state, at least for small systems ($N \leq 15$).   (4) Certain wavefunctions from this family, albeit the nonunitary members of the family, have very high overlap with the ground state of potentially experimentally relevant Hamiltonians.

\acknowledgments

The authors acknowledge helpful discussions with Nick Read, Paul Fendley, Joost Slingerland, and Victor Gurarie.   E.~R. is supported by DOE grants DE-FG03-02ER-45981 and DE-SC0002140.


\begin{thebibliography}{50}
\bibitem{endnotespectrum} It is likely that additional terms need to be added to $H$ to obtain the correct quasihole spectrum corresponding to a given RCFT.
\end{thebibliography}
\end{document}